\documentclass[twocolumn,preprintnumbers,amsmath,amssymb]{revtex4}
\usepackage{graphicx}% Include figure files
\usepackage{dcolumn}% Align table columns on decimal point
\usepackage{subfigure}% for two column figures
\usepackage{bm}% bold math

\newcommand{\ket}[1]{\left|#1\right\rangle}

\begin{document}

%\preprint{APS/123-QED}

\title{Decoherence in Two-Dimensional Quantum Random Walks with Traps}% Force line breaks with \\

\author{Meltem G\"{o}n\"{u}lol$^1$}
\author{Ekrem Aydiner$^2$}
\email{ekrem.aydiner@istanbul.edu.tr}
\author{\"{O}zg\"{u}r E. M\"{u}stecapl{\i}o\~{g}lu$^3$}
\affiliation{$^1$Department of Physics, Dokuz Eyl\"{u}l University,
35160, Izmir, Turkey \\ $^2$Department of Physics,
Istanbul University, 34134, Istanbul, Turkey \\
$^3$Department of Physics, Ko\c{c} University, Sariyer, 34450, Istanbul,
Turkey}

%\date{December 22, 2008}

\begin{abstract}
Quantum random walk in a two-dimensional lattice with randomly
distributed traps is investigated. Distributions of quantum walkers
are evaluated dynamically for the cases of Hadamard, Fourier, and
Grover coins, and quantum to classical transition is examined as a
function of the density of the traps. It is shown that traps act as
a serious and additional source of quantum decoherence. Furthermore,
non-trivial temporal dependence of the standard deviation of the
probability distribution of the walker is found when the trapping
imperfections are introduced.
\end{abstract}

%\pacs{03.67.Lx,05.40.Fb,03.65.Yz}
% PACS, the Physics and Astronomy classification Scheme.



\maketitle

%%%%%%%%%%%%%%%%%%%%%%%%%%%%%%%%%%%%%%%%%%%%%%%%%%%%%%%%%%%%%%%%%%%%%%%%%%%%%%%%%%
%%%%%%%%%%%%%%%%%%%%%%%%%%%%%%%%%%%%%%%%%%%%%%%%%%%%%%%%%%%%%%%%%%%%%%%%%%%%%%%%%%
%%%%%%%%%%%%%%%%%%%%%%%%%%%%%%%%%%%%%%%%%%%%%%%%%%%%%%%%%%%%%%%%%%%%%%%%%%%%%%%%%%

\section{\label{sec:level1}Introduction}

A simple stochastic process can be introduced by the motion of a
particle that can move in certain directions with some probabilities
such that progress of the particle is independent of its preceding
movements. Finding the spatial probability distribution of the
particle, starting such a randomized motion from a given location,
is the statement of the so-called random walk problem. Random walk
and its more generalized extensions have been used as simple models
for many physical systems, in particular in solid-state physics and
in astronomy \cite{chandra}, and in polymer models \cite{barber}.
Classical random walks have found variety of applications in other
fields, such as in economics and in computational sciences, as well.
In particular, powerful randomized algorithms, especially for graph
connectivity, satisfiability, probability amplification,
\cite{motwani} and Markovian chain simulations \cite{jerrum}, have
been developed based on classical random walks.

Quantum computation and quantum information are one of the most
active research fields nowadays \cite{nielsen}. A quantum computer
would be capable to run quantum algorithms, such as Deutsch-Jozsa
\cite{deutsch}, Shor's \cite{shor} or Grover's algorithm
\cite{grover}, much faster than any classical algorithms running on
classical computers. One direction of research aiming to develop
more quantum algorithms is to implement some classical algorithms
directly by considering their quantum analogs. In this respect,
quantum analogs of random walks, the so-called the quantum random
walks (QRWs) have received much attention recently (see. e.g., Refs.
\cite{kempe,Konno,Andraca}). Some new quantum algorithms based on
QRWs have already been proposed
\cite{childs,shenvi,goldstone,ambainis}. It is proven that a
discrete time quantum walk can be used for searching unsorted
database with a quadratic speedup \cite{shenvi}, while the
continuous time quantum walks can be exploited for traversing
certain graphs exponentially faster than any classical algorithm
\cite{childs}.

There are various schemes and systems including ion traps
\cite{travaglione}, optical lattices \cite{dur,eckert}, cavity QED
\cite{sanders}, optical cavity \cite{knight}, and linear optics
\cite{do,pathak} that have been considered for practical realization
of QRWs. Using nuclear-magnetic resonance two- and three-qubit
quantum information processors, QRW has been demonstrated on a
square for its continuous time \cite{du} and discrete time
\cite{ryan} versions. Such direct implementations of QRW are not
essential for quantum algorithm developments. As long as there
exists a quantum computer, QRW and associated quantum algorithm can
always be realized on it. On the other hand, QRWs provide deep
insight into transition to classical behavior out of a quantum
behavior as well as into decoherence and coherent control of quantum
systems. Rich quantum dynamics available within QRWs makes them
appealing systems \emph{per se}. A well-known property of a quantum
walker is that it is more de-localized than a classical walker such
that a quantum walker spreads out quadratically faster in time than
a classical walker.

Quantum walkers are highly delocalized particles, and their QRW is
very sensitive to decoherence. The rapid transition of QRW to a
classical walk may limit their implementations and applications, yet
at the same time, small amount of decoherence can be beneficial to
speed up quantum algorithms based on QRWs \cite{kendon}. In order to
develop optimized quantum walks on noisy quantum channels, it is
necessary to investigate decoherence mechanisms and characterize
their effects on the propagation of a quantum walker. Decoherence in
QRW can be discussed in general terms (for a review, see Ref.
\cite{kendon2006}) as well as in more specialized contexts,
depending on a particular implementation. For that aim, QRWs on a
line under various decoherence mechanisms, such as random
measurements or broken links \cite{romanelli}, and under various
environmental noises, such as phase flip, bit flip, and generalized
amplitude damping channels \cite{chandra2007}, have been examined.
The general purpose of these works is to understand and resolve the
different decoherence problems to be encountered in various
one-dimensional implementations of QRWs. In this work we contribute
the same purpose by considering an additional decoherence mechanism
and by focusing on different QRWs in two dimensions. Keeping the
walkers on the lines or planes of interest can be nontrivial and
practical realizations of QRWs may face a serious problem of loss of
walkers during the evolution. In this work, we consider loss of
walkers as a decoherence mechanism.

A natural imperfection that can cause nonunitary evolution of a
quantum walker would be loss of the walker at certain places along
the propagation directions. This may correspond to absorption of
photon in optical implementations, loss of atom due to thermal
fluctuations or collisions in atom traps or in optical lattices, or
immobilization of quasi-particles in solids and so on. Indeed, in
the case of classical RW problem, the possibility of the trapping of
classical random walks by randomly distributed traps on a lattice
has been extensively studied using both theoretical and Monte Carlo
methods
\cite{Havlin1,Havlin2,Grassberger,Kayser,Stanley,Redner,Prasad}. The
exact source of the loss of the quantum walker would depend on a
particular implementation. In certain systems, such as optical
lattices, imperfections can be introduced on purpose and
controllably such as by adjusting the depths of the atom traps or by
optical addressing of a set of lattice sites. The results to be
reported in this paper are a systematic characterization of effects
of immobilizing centers on the QRW and are independent of a specific
implementation.

QRW in higher dimensions, especially in two dimensions, have been
studied extensively \cite{Watabe,Inui}. In higher dimensions,
decoherence effect can be of different significance than in one
dimension \cite{Mackay,oliveira}. In this paper, we numerically
investigate the effect of decoherence that causes quantum to
classical transition in a two-dimensional (2D) random walk with
traps. Typical quantum walks, namely, Hadamard (H), Fourier (F) and
Grover (G) walks are considered.  Their relative endurances against
decoherence are characterized. Traps along the paths of the walkers
are recognized as a serious source of decoherence that should be
reckoned with for a potential implementation of the quantum walk, in
addition to broken links or random measurements.

The paper is organized as follows. We first present a short review
of the theory of the two-dimensional QRW together with our model of
walk on a lattice with traps in Sec. \ref{sec:model}. Results of the
numerical simulations and their discussions are given in Sec.
\ref{sec:results}. Finally, in Sec. \ref{sec:conclusion}, we
summarize the conclusions of this work.
%%%%%%%%%%%%%%%%%%%%%%%%%%%%%%%%%%%%%%%%%%%%%%%%%%%%%%%%%%%%%%%%%%%%%%%%%%%%%%%%%%
%%%%%%%%%%%%%%%%%%%%%%%%%%%%%%%%%%%%%%%%%%%%%%%%%%%%%%%%%%%%%%%%%%%%%%%%%%%%%%%%%%
%%%%%%%%%%%%%%%%%%%%%%%%%%%%%%%%%%%%%%%%%%%%%%%%%%%%%%%%%%%%%%%%%%%%%%%%%%%%%%%%%%
\section{\label{sec:model} Quantum Random Walks in a 2D lattice with traps}

We consider a QRW in a two-dimensional lattice with
traps. The traps at a given density are randomly distributed before
the walk is started then the trap locations remain frozen. For a QRW
on an infinite two-dimensional lattice, the coin space is
$\mathcal{H}_{4}$, spanned by the basis states $\{\ket{j,k},j,k\in
\{0,1\}\}$, and the position space is $\mathcal{H}_{\infty}$,
spanned by the basis states $\{\ket{m,n},m,n$ integers\}.  The state
of a quantum walker at time $t$ in
$\mathcal{H}_{4}\otimes\mathcal{H}_{\infty}$ space is given by
\begin{equation}\label{state}
\ket{\psi(t)}=\sum_{j,k=0}^{1}\sum_{n,m=-\infty}^{\infty}A_{jkmn}(t)\ket{j,k}\ket{m,n}.
\end{equation}

One step of the quantum walker is defined by the unitary operator
$U$ such that
\begin{equation}\label{}
U=S(I_{4}\otimes C),
\end{equation}
where $I_{4}$ is the $4\times 4$ identity matrix, $C$ is the coin
operator which written as
\begin{equation}\label{}
C=\sum_{j,k=0}^{1}\sum_{j',k'=0}^{1}C_{jkj'k'}\ket{j,k}\ket{j',k'},
\end{equation}
and $S$ is the shift operator described by
\begin{equation}\label{}
S\ket{j,k}\ket{m,n}=\ket{j,k}\ket{m+(-1)^{j},n+(-1)^{k}}.
\end{equation}

We analyze the results for three different coins which are Hadamard,
Fourier, and Grover. H, F, and G coins in two dimensions are,
respectively, given by
\begin{eqnarray}
C_H&=& \frac{1}{2}\left(
        \begin{array}{cccc}
          1 & 1 & 1 & 1 \\
          1 & -1 & 1& -1\\
          1 & 1 & -1 & -1\\
          1 & -1 & -1 & 1\\
        \end{array}
      \right), \\
C_F&=& \frac{1}{2}\left(
        \begin{array}{cccc}
          1 & 1 & 1 & 1 \\
          1 & i & -1& -i\\
          1 & -1 & 1 & -1\\
          1 & -i & -1 & i\\
        \end{array}
      \right), \\
C_G&=& \frac{1}{2}\left(
        \begin{array}{cccc}
          -1 & 1 & 1 & 1 \\
          1 & -1 & 1& 1\\
          1 & 1 & -1 & 1\\
          1 & 1 & 1 & -1\\
        \end{array}
      \right).
\end{eqnarray}

The quantum walker begin to walk at $(0,0)$ point in the two-
dimensional lattice. The initial states for Hadamard, Fourier, and
Grover walks were chosen to guarantee a maximum spreading when the
walk starts at the origin \cite{Tregenna}, and as such for the H,
the F, and the G coins, they are, respectively, taken to be as
\begin{eqnarray}\label{psi_sim_hadamard}
\ket{\psi(0)}_H&=&\frac{1}{2}(\ket{00}+i\ket{01}-i\ket{10}+\ket{11})\ket{0,0},\\
\ket{\psi(0)}_F&=&\frac{1}{2}((\ket{00}+\frac{1-i}{\sqrt{2}}\ket{01}
+\ket{10}\nonumber\\
&-&\frac{1-i}{\sqrt{2}}\ket{11})\ket{0,0}),\\
\ket{\psi(0)}_G&=&\frac{1}{2}((\ket{00}-\ket{01}-\ket{10}+\ket{11})\ket{0,0}).
\end{eqnarray}
\begin{figure}[htbp]
\includegraphics[width=3 in]{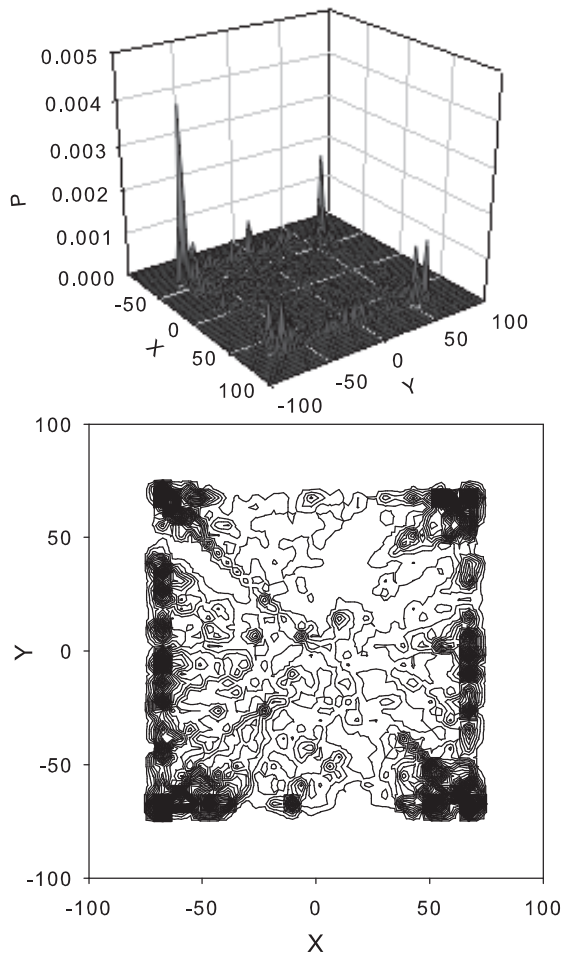}
\caption{{\label{Hadamard_sim1}} The probability distribution of the
Hadamard walk for $p=0.01$ after 100 iterations.}
\end{figure}
\begin{figure}[htbp]
\includegraphics[width=3 in]{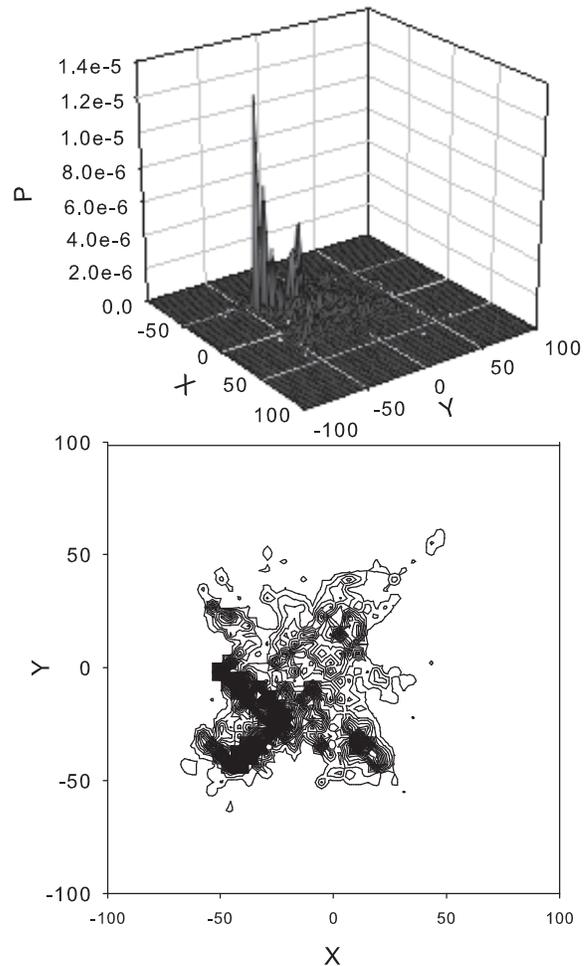}
\caption{{\label{Hadamard_sim2}} The probability distribution of the
Hadamard walk for $p=0.1$ after 100 iterations.}
\end{figure}

Applying $U$ on the state of the walker (\ref{state}), the relation
that determines the spatiotemporal evolution of the quantum walker
is obtained to be as
\begin{equation}\label{genlik}
A_{jkmn}(t+1)=\sum_{j',k'=0}^{1}C_{jkj'k'}A_{j'k'm_jn_j}(t),
\end{equation}
with $m_j=m-(-1)^{j}$ and $n_j=n-(-1)^{k}$. With the amplitude of
new positions of the quantum walker given by Eq.(\ref{genlik}), the
probability distribution of the walker at position $\ket{m,n}$ at
time $t$ is determined by
\begin{equation}\label{olasilik}
P_{m,n}(t)=\sum_{j,k=0}^{1}|A_{j,k;m,n}(t)|^2.
\end{equation}

In the calculations, we assume that the traps are completely
absorbing, so that if the quantum walker falls into a trap, it is
annihilated with vanishing probability amplitudes at any later time.
In the next section, the results of probability distributions
obtained by numerical simulations for the cases of Hadamard,
Fourier, and Grover walks with different trap densities will be
presented. In particular, the examination of the standard deviation
$\sigma$ will be used to characterize quantum to classical
transition of the walks. In the case of a classical walk,
$\sigma\propto\sqrt{t}$, while a quadratic gain in the spread is
achieved in a quantum walk for which $\sigma\propto t$ due to
quantum coherence. It is known that unitary noises such as broken
links or nonunitary disturbances such as random measurements destroy
this gain and make the quantum walk a classical one. Here, we
explore how the traps, causing nonunitary evolution of the quantum
walker in the two-dimensional lattice, might suppress the benefit of
the quantum walk. As the traps are randomly distributed, it is
necessary to consider many different initial configurations of the
traps, so that we calculate
\begin{equation}\label{}
\langle\sigma\rangle=\frac{1}{M}\sum_{r=1}^{M}\sigma_{r},
\end{equation}
where $M$ denotes the number of different configurations of random
traps.
%%%%%%%%%%%%%%%%%%%%%%%%%%%%%%%%%%%%%%%%%%%%%%%%%%%%%%%%%%%%%%%%%%
%%%%%%%%%%%%%%%%%%%%%%%%%%%%%%%%%%%%%%%%%%%%%%%%%%%%%%%%%%%%%%%%%%%%%%%%%%
%%%%%%%%%%%%%%%%%%%%%%%%%%%%%%%%%%%%%%%%%%%%%%%%%%%%%%%%%%%%%%%%%%%%%%%%%%%%%
\section{\label{sec:results}Results and Discussions}

In our simulations we have used an ensemble of 250 different
configurations of randomly placed traps and the ensemble averaged
results are presented in the figures for the calculated variance. We
first investigate the case of a Hadamard walk. Quantum walker starts
at the origin and its walk is numerically simulated for various trap
densities. Figure \ref{Hadamard_sim1} shows the probability
distribution of the walker at time $t=100$ for a trap density of
$p=0.01$. The walk is seen to be still retaining its quantum
character. One would expect a symmetric centrally peaked
Gaussian-like distribution about the origin in the classical walk.
We have found that such a behavior emerges and the walk becomes
classical for about $p=0.1$ at $t=100$, which is shown in Fig.
\ref{Hadamard_sim2}.
\begin{figure}[tbp]
\includegraphics[width=3.5in]{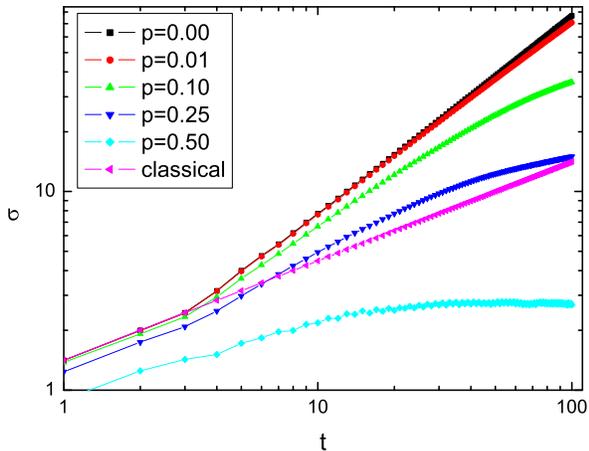}
\caption{\label{Hadamard_standart sapma} (Color online) Time
dependence of the standard deviation of the Hadamard walk in 2D
trapping lattice with trap densities $p=0$, $p=0.01$, $p=0.1$,
$p=0.25$, $p=0.5$, and classical random walk in 2D lattice.}
\end{figure}
\begin{figure}[h!]
\includegraphics[width=3.3 in]{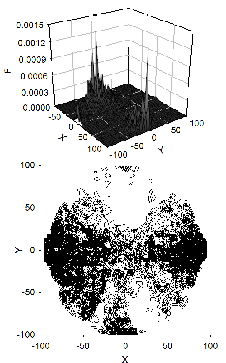}
\caption{{\label{Fourier_sim1}} The probability distribution of the
Fourier walk for $p=0.01$ after 100 iterations.}
\end{figure}
\begin{figure}[htbp]
\includegraphics[width=3 in]{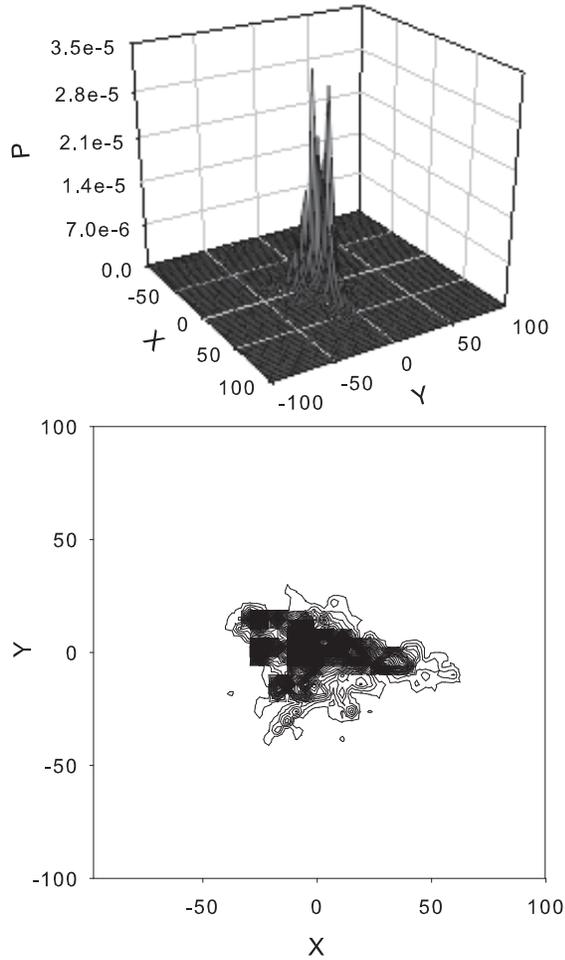}
\caption{{\label{Fourier_sim2}} The probability distribution of the
Fourier walk for $p=0.1$ after 100 iterations.}
\end{figure}

The time dependence of the standard deviation for the Hadamard walk
for different trap densities is plotted in Fig.
\ref{Hadamard_standart sapma} with a log-log scale. When there are
no traps, $p=0$, the highest spatial spread of the walkers or the
most quantum behavior is available. When trap densities increase,
the standard deviation is dramatically altered. For small amount of
traps, the spatial spread is gradually decreased but still $\sigma$
remains larger than the classical one, $\sigma_{cl}$, and grows
faster. However, even when a modest amount of traps are introduced,
the temporal behavior is no longer linear and $\sigma$ can grow as
slow as - or even slower - than $\sigma_{cl}$. This is in stark
contrast with the other types of decoherence sources, such as broken
links \cite{oliveira}, for which simple power laws describe the time
dependence of $\sigma$. The emergence of the nonlinear behavior of
$\sigma$ in a trapped lattice can be attributed to the fact that the
walkers can survive in a trapped lattice only for a time and their
probability of survival is related to the Kohlrausch-Williams-Watts
functions (stretched exponentials) of the form
$\exp{(-t/\tau)^{\beta}}$ with $0<\beta<1$ as the stretching
exponent and $\tau$ as a characteristic relaxation time
\cite{kohlrausch,williams,phillips}.

When most of the lattice is trapped randomly, so that $p>0.5$, the
walkers become highly localized with $\sigma$ being significantly
smaller than $\sigma_{cl}$. In other sources of decoherence, when
random events are as frequent, $\sigma$ can become less than
$\sigma_{cl}$, too \cite{oliveira}. However, traps force a more
severe reduction which increases in time. Although $\sigma$ can be
smaller in value than the classical one, it cannot grow slower than
the classical walk in broken link disturbances. With the traps,
however, both the magnitude of $\sigma$ and its growth rate can be
made smaller than those of the classical walk. By examining the
slopes of the curves in Fig. \ref{Hadamard_standart sapma}, it can
be deduced that the transition from the quantum to the classical
behavior happens at a time $\tau_{decoh}\approx 5/p$, which is the
decoherence time. After that, the growth rate of $\sigma$ gradually
becomes smaller than the classical one, which is a reflection of the
relatively short survival time and a stretched exponential behavior
of the survival probability. In the broken links caused decoherence
$\tau_{decoh}\approx 3/p$ is found \cite{oliveira}. This suggests
that two-dimensional lattices can endure only slightly longer to
decoherence due to traps.

Let us now look at the other typical quantum walks.  The probability
distribution of the walker in the Fourier walk when the trap density
is $p=0.01$ is given in Fig. \ref{Fourier_sim1} at iteration
$t=100$. Analogous to the Hadamard walk, a significant quantum
behavior is maintained at such low trap densities, while as shown in
Fig. \ref{Fourier_sim2} for $p=0.1$ and $t=100$ the classical
behavior is emerged.
\begin{figure}[htbp]
\includegraphics[width=3.5 in]{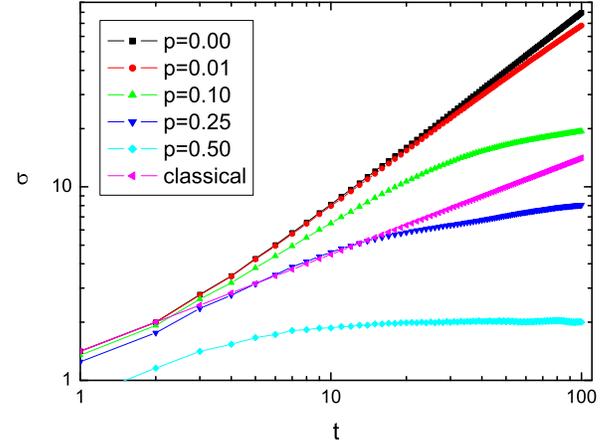}
\caption{\label{Fourier_standart sapma} (Color online) Time
dependence of the standard deviation of the Fourier walk in 2D
trapping lattice with trap densities $p=0$, $p=0.01$, $p=0.1$,
$p=0.25$, $p=0.5$ and classical random walk in 2D lattice.}
\end{figure}
\begin{figure}[htbp]
\includegraphics[width=3.05 in]{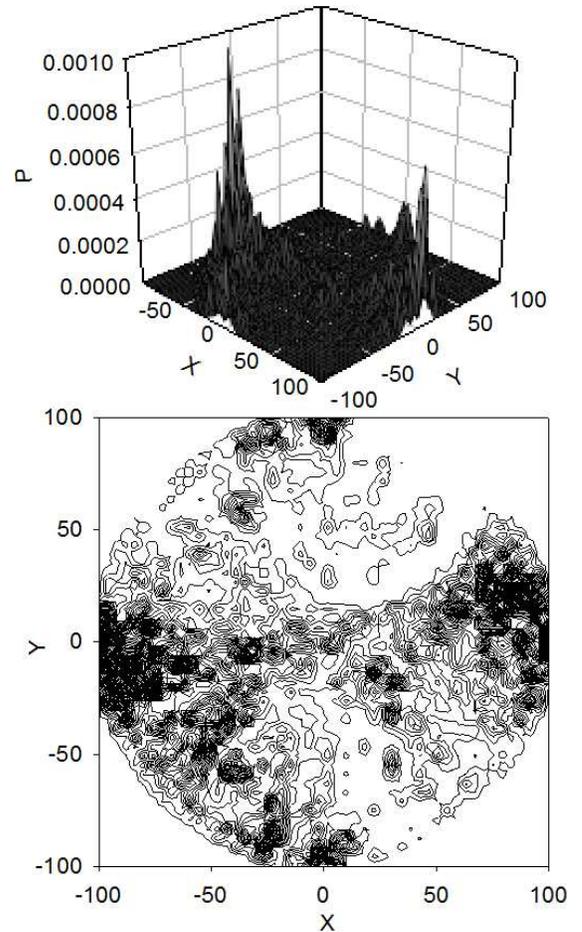}
\caption{{\label{Grover_sim1}} The probability distribution of the
Grover walk for $p=0.01$ after 100 iterations.}
\end{figure}
The time evolution of the standard deviation for the Fourier walk
for different trap densities is depicted in Fig.
\ref{Fourier_standart sapma} with a log-log scale. A strong
nonlinear behavior is found in this case as in the Hadamard walk.
When Fig. \ref{Fourier_standart sapma} is compared to Fig.
\ref{Hadamard_standart sapma}, it is seen that the Fourier walk
becomes classical more quickly than the Hadamard walk. This
observation is in agreement with the broken links case
\cite{oliveira}.
\begin{figure}[htbp]
\includegraphics[width=3 in]{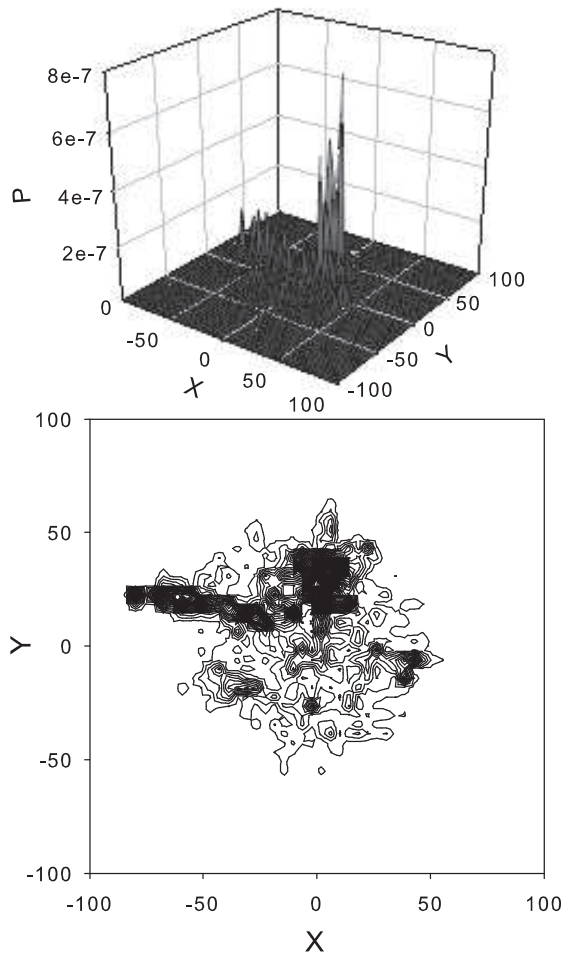}
\caption{{\label{Grover_sim2}} The probability distribution of the
Grover walk for $p=0.1$ after 100 iterations.}
\end{figure}

\begin{figure}[htbp]
\includegraphics[width=3.5 in]{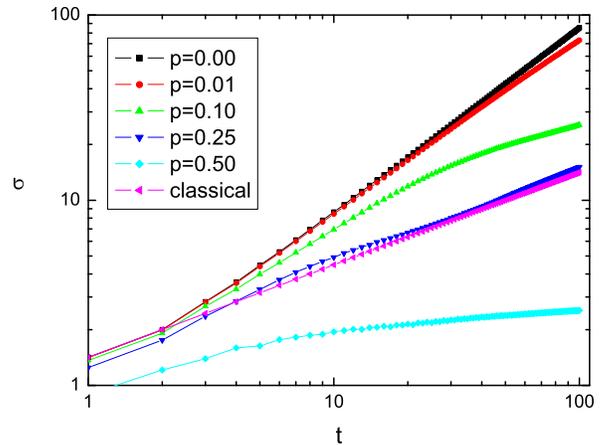}
\caption{\label{Grover_standart sapma} (Color online) Time
dependence of the standard deviation of the Grover walk in 2D
trapping lattice with trap densities $p=0$, $p=0.01$, $p=0.1$,
$p=0.25$, $p=0.5$ and classical random walk in 2D lattice.}
\end{figure}

Finally, we look at the case of the Grover walk for which Fig.
\ref{Grover_sim1} describes the probability distribution at time
$t=100$ for the case of a trap density $p=0.01$, and Fig.
\ref{Grover_sim2} for $p=0.1$. The low and the high trap density
behaviors lead to the similar conclusions as in the Grover and the
Fourier walks, while the time dependence of the standard deviation
in Fig. \ref{Grover_standart sapma} indicates that the Hadamard walk
comes to the classical behavior more slowly than the Fourier and the
Grover walks. The Hadamard walk in two-dimensional lattice can be
said to be more durable in decoherence due to traps while the
Fourier is the least durable. This observation is also valid for the
case of decoherence due to broken links \cite{oliveira}, which is in
fact a fundamentally based on symmetry properties of the probability
distributions \cite{oliveira} when there is no decoherence and thus
should be valid for any type of imperfections introduced.

%%%%%%%%%%%%%%%%%%%%%%%%%%%%%%%%%%%%%%%%%%%%%%%%%%%%%%%%%%%%%%%%%%
%%%%%%%%%%%%%%%%%%%%%%%%%%%%%%%%%%%%%%%%%%%%%%%%%%%%%%%%%%%%%%%%%%%%%%%%%%
%%%%%%%%%%%%%%%%%%%%%%%%%%%%%%%%%%%%%%%%%%%%%%%%%%%%%%%%%%%%%%%%%%%%%%%%%%%%%

\section{\label{sec:conclusion}Conclusions}

Introducing traps along the path of a random walker has been
investigated in detail classically in the context of the survival
probability of particles in solid-state systems with defects and
impurities. We examine this problem for the case of quantum random
walk and show that traps act as a significant source of quantum
decoherence and make the walk a classical one even when a modest
number of traps are present. We consider the cases of three
different quantum coins, which are Hadamard, Fourier, and Grover.
The corresponding initial states for each coin are chosen to
guarantee a maximum spreading when the walk starts at the origin.
Characterization of the traps as a source of decoherence reveals
that trapping forces transition to a classical walk from a quantum
walk in a relatively longer time in comparison to the broken links.
While all three types of the quantum walks become classical after a
time decreasing with the trap density, it is found that quantum walk
with a Hadamard coin is found to be the most tolerant to the density
of the traps. A quantum walker with a Fourier coin would exhibit the
lowest endurance against traps. Major difference in a trapped
lattice is the survival time of the walkers that causes slower
growth rates of the standard deviations than classical walks and
lack of simple power-law temporal growth of the standard deviations
due to the stretched exponential behavior of the survival
probabilities of the walkers. In addition to broken links and random
measurements, traps corresponding to defects and impurities thus
should be considered as a serious and additional source of
decoherence in practical implementations of quantum random walks.
%%%%%%%%%%%%%%%%%%%%%%%%%%%%%%%%%%%%%%%%%%%%%%%%%%%%%%%%%%%%%%%%%%
%%%%%%%%%%%%%%%%%%%%%%%%%%%%%%%%%%%%%%%%%%%%%%%%%%%%%%%%%%%%%%%%%%%%%%%%%%
%%%%%%%%%%%%%%%%%%%%%%%%%%%%%%%%%%%%%%%%%%%%%%%%%%%%%%%%%%%%%%%%%%%%%%%%%%%%%
\section{\label{sec:acknowledgement }Acknowledgements }

M.G. acknowledges the travel support by DPT. \"{O}.E.M. gratefully
acknowledges the hospitality and the support by the Dokuz Eyl\"{u}l
University.


\newpage

\begin{thebibliography}{99}

\bibitem{chandra} S. Chandrasekhar, Rev. Mod. Phys. {\bf 15}, 1
(1943).

\bibitem{barber} M. N. Barber and B. W. Ninham, {\it Random and
Restricted Walks: Theory and Applications} (Gordon and Breach, New
York, 1970).

\bibitem{motwani} R. Motwani and P. Raghavan, {\it Randomized Algorithms} (Cambridge
University Press, Cambridge, U.K., 1995).

\bibitem{jerrum} M. Jerrum and A. Sinclair, in {\it Approximation Algorithm for
NP-HARD Problems}, edited by D. S. Hochbaum (PWS Publishing, Boston,
1996), Chap. 12, pp. 482–-520.

\bibitem{nielsen} M. A. Nielsen and I. L. Chuang, {\it Quantum
Computation and Quantum Information} (Cambridge University Press,
Cambridge, England, 2000).

\bibitem{deutsch} D. Deutsch and R. Jozsa, Proc. R. Soc. London Ser. A
{\bf 439}, 553 (1992).

\bibitem{shor} P. W. Shor, in {\it Proceedings of the 35th Annual
Symposium on the Foundations of Computer Science}, edited by S.
Goldwasser (IEEE Computer Society Press, Los Alamitos, CA, 1994),
pp. 124--134.

\bibitem{grover} L. K. Grover, Phys. Rev. Lett. {\bf 79}, 325
(1997).

\bibitem{kempe} J. Kempe, Contemp. Phys. {\bf 44}, 307 (2003).

\bibitem{Andraca} S. E. Venegas-Andraca,  {\it Quantum Walks for Computer
Scientists} (Morgan and Claypool Publishers, San Rafael, CA, 2008).

\bibitem{Konno} N. Konno, in {\it Quantum Potential Theory}, Lecture Notes in Mathematics Vol.
1954, edited by U. Franz and M. Sch\"urmann (Springer-Verlag,
Heidelberg 2008), pp. 309-452.

\bibitem{childs} A.M. Childs, R. Cleve, E. Deotto, E. Farhi, S. Gutmann, and
D.A. Spielman, {\it Proceedings of the 35th ACM Symposium on Theory
of Computing} (ACM Press, New York, 2003), p. 59.

\bibitem{shenvi} N. Shenvi, J. Kempe, and K. Birgitta Whaley, Phys. Rev. A {\bf 67},
052307 (2003).

\bibitem{goldstone} A. M. Childs and J. Goldstone, Phys. Rev. A {\bf 70}, 022314 (2004).

\bibitem{ambainis} Andris Ambainis, Julia Kempe, Alexander Rivosh, {\it  Proceedings of the 16th ACM-SIAM
SODA} (Society for Industrial and Applied Mathematics, Philadelphia,
PA, 2005), p. 1099.

\bibitem{travaglione} B. C. Travaglione and G. J. Milburn, Phys. Rev. A {\bf 65}, 032310 (2002).

\bibitem{dur} W. Dur, R. Raussendorf, V. M. Kendon, and H. J. Briegel, Phys.
Rev. A {\bf 66}, 052319 (2002).

\bibitem{eckert} K. Eckert, J. Mompart, G. Birkl, and M. Lewenstein, Phys. Rev. A
{\bf 72}, 012327 (2005).

\bibitem{sanders} B. C. Sanders, S. D. Bartlett, B. Tregenna, and P. L. Knight,
Phys. Rev. A {\bf 67}, 042305 (2003).

\bibitem{knight} P. L. Knight, E. Roldán, and J. E. Sipe, Opt. Commun. {\bf 227}, 147
(2003).

\bibitem{do} B. Do, M. L. Stohler, S. Balasubramanian, D. S. Elliott, C.
Eash, E. Fischbach, M. A. Fischbach, A. Mills, and B. Zwickl, J.
Opt. Soc. Am. B {\bf 22}, 499 (2005).

\bibitem{pathak}  P. K. Pathak and G. S. Agarwal, Phys. Rev. A {\bf 75}, 032351
(2007).

\bibitem{du} J. Du, H. Li, X. Xu, M. Shi, J. Wu, X. Zhou, and R. Han, Phys.
Rev. A {\bf 67}, 042316 (2003).

\bibitem{ryan} C. A. Ryan, M. Laforest, J. C. Boileau, and R. Laflamme, Phys.
Rev. A {\bf 72}, 062317 (2005).

\bibitem{kendon} V. Kendon and B. Tregenna, Phys. Rev. A {\bf 67}, 042315
(2003).

\bibitem{kendon2006} V. Kendon, Math. Struct. Comp. Sc.
{\bf 17}, 1169 (2007).

\bibitem{romanelli} A. Romanelli, R. Siri, G. Abal, A. Auyuanet and R. Donangelo,
Physica A {\bf 347}, 137 (2005). %Phys. A = Physica A

\bibitem{chandra2007} C. M. Chandrashekar, R. Srikanth, and Subhashish
Banerjee, Phys. Rev. A {\bf 76}, 022316 (2007).

\bibitem{Havlin1} S. Havlin, G. H. Weiss, J. E. Kiefer and M. Dishon,
J. Phys. A  \textbf{17}, L347 (1984).

\bibitem{Havlin2} S. Havlin, M. Dishon, J. E. Kiefer, and G. H.
Weiss, Phys. Rev. Lett. \textbf{53}, 407 (1984).

\bibitem{Grassberger} P. Grassberger and I. Procaccia, J. Chem.
Phys. \textbf{77}, 6281 (1982).

\bibitem{Kayser} R. F. Kayser and J. B. Hubbard, Phys. Rev. Lett. \textbf{51}, 79 (1983).

\bibitem{Stanley} H. E. Stanley, K. Kang, S. Redner, and R. L. Blumberg, Phys. Rev. Lett. \textbf{51}, 1223 (1983).

\bibitem{Redner} S. Redner and  K. Kang, Phys. Rev. Lett. \textbf{51}, 1729 (1983).

\bibitem{Prasad} M. A. Prasad and M. Nagarajan, J. Phys. A \textbf{32}, 7665 (1999).

\bibitem{Watabe} K. Watabe, N. Kobayashi, M. Katori and N. Konno, Phys. Rev. A
{\bf 77}, 062331 (2008).

\bibitem{Inui} N. Inui, Y. Konishi, and N. Konno, Phys. Rev. A {\bf 69}, 052323 (2004).

\bibitem{Mackay} T. D. Mackay, S. D. Bartlett, L. T. Stephenson, and B.
C. Sanders, J. Phys. A \textbf{35}, 2745 (2002)

\bibitem{oliveira} A. C. Oliveira, R. Portugal, and R. Donangelo, Phys. Rev. A
\textbf{74}, 012312 (2006)

\bibitem{Tregenna}  B. Tregenna, W. Flanagan, R. Maile, and V. Kendon,
New J. Phys. {\bf5}, 83 (2003).

\bibitem{kohlrausch} R. Kohlrausch, Pogg. Ann. Phys. Chem. {\bf 91}, 179 (1854).

\bibitem{williams} G. Williams and D. Watts, Trans. Faraday Soc. {\bf 66}, 80 (1970).

\bibitem{phillips} J. C. Phillips, Rep. Prog. Phys. {\bf 59}, 1133 (1996).

\end{thebibliography}
\end{document}